# Room-temperature van der Waals 2D ferromagnet switching by spin-orbit torques

Weihao Li, [1,2,†] Wenkai Zhu,[1,2,†] Gaojie Zhang,[3,4] Hao Wu,[3,4] Shouguo Zhu,[1,2] Runze Li,[1,2] Enze Zhang,[1,2] Xiaomin Zhang,[1,2] Yongcheng Deng,[1,2] Jing Zhang,[1] Lixia Zhao,[1,5] Haixin Chang,[3,4,*] and Kaiyou Wang[1,2,*]

[1]State Key Laboratory of Superlattices and Microstructures, Institute of Semiconductors, Chinese Academy of Sciences, Beijing 100083, China
[2]Center of Materials Science and Optoelectronics Engineering, University of Chinese Academy of Sciences, Beijing 100049, China
[3]Center for Joining and Electronic Packaging, State Key Laboratory of Material Processing and Die & Mold Technology, School of Materials Science and Engineering, Huazhong University of Science and Technology, Wuhan 430074, China
[4]Wuhan National High Magnetic Field Center, Huazhong University of Science and Technology, Wuhan 430074, China
[5]School of Electrical and Electronic Engineering, Tiangong University, Tianjin 300387, China
†These authors contributed equally to this work.
*Corresponding author. E-mail: hxchang@hust.edu.cn; kywang@semi.ac.cn

**Keywords:**

Room-temperature, van der Waals 2D ferromagnet, $Fe_3GaTe_2$, spin-orbit torque

**Abstract:**

Emerging wide varieties of the two-dimensional (2D) van der Waals (vdW) magnets with atomically thin and smooth interfaces holds great promise for next-generation spintronic devices. However, due to the lower Curie temperature of the vdW 2D ferromagnets than room temperature, electrically manipulating its magnetization at room temperature has not been realized. In this work, we demonstrate the perpendicular magnetization of 2D vdW ferromagnet $Fe_3GaTe_2$ can be effectively switched at room temperature in $Fe_3GaTe_2$/Pt bilayer by spin-orbit torques (SOTs) with a relatively low current density of $1.3\times10^7 A/cm^2$. Moreover, the high SOT efficiency of $\xi_{DL}{\sim}0.22$ is quantitatively determined by harmonic measurements, which is higher than those in Pt-based heavy metal/conventional ferromagnet devices. Our findings of room-temperature vdW 2D ferromagnet switching by SOTs provide a significant basis for the development of vdW-ferromagnet-based spintronic applications.

**1. Introduction**

Spin-orbit torque (SOT) based magnetic random-access memory (MRAM) is a leading candidate for nonvolatile spin-memory and logic applications.[1–8] Heterojunctions based on heavy metal (HM)/three-dimensional (3D) ferromagnet (FM) with perpendicular magnetic anisotropy (PMA) have been extensively studied and utilized.[2,9] However, the use of 3D perpendicular ferromagnets in such structures limits the reduction of device size since the interface-induced PMA would sharply be weakened by thermal effect during device miniaturization.[10] Fortunately, the recent discovery of strong perpendicular two-dimensional ferromagnets (2DFMs), such as $Fe_3GeTe_2$[11,12] and $Fe_3GaTe_2$,[13] which possess atomically flat surfaces and exhibit magnetic properties at their 2D limit thickness, have potential to break through these limitations and offer a promising platform to build up advanced spintronic devices. Among them, SOT-driven magnetization switching in vdW ferromagnet is highly pursued for the practical applications due to its energy-efficient and ultrafast advantages. Although SOT switching has been demonstrated in vdW ferromagnets such as $Fe_3GeTe_2$[14–17] and $Cr_2Ge_2Te_6$ (CGT)[18–20], all these devices can not satisfy the room-temperature application requirements.

The intrinsic above-room-temperature ferromagnetic vdW $Fe_3GaTe_2$ is a very competitive ferromagnet due to its robust PMA with Curie temperature $T_c \sim$ 350-380 K,[13] which are critical for the magnetoelectronic devices such as MTJ[21–24] and MRAM. Typically, a large tunnel magnetoresistance of 85% at 300 K was recently demonstrated in $Fe_3GaTe_2$/$WSe_2$/$Fe_3GaTe_2$ MTJs.[21] However, room-temperature SOT devices based on vdW FM, acting as an essential building block for SOT-MARM, have not yet been realized.

In this work, we utilize a multi-layer $Fe_3GaTe_2$ as the magnetic layer and deposit 6 nm Pt onto it to fabricate the bilayer Hall device. With the assistance of an external in-plane magnetic field, the perpendicular magnetization switching of $Fe_3GaTe_2$ with a low current density of $1.3\times10^7$ A/cm$^2$ has been achieved at room temperature (300 K). The low switching current density is attributed to the high dampling-like SOT efficiency in the $Fe_3GaTe_2$/Pt bilayer, which is quantitatively determined to be $\xi_{DL}\sim$ 0.22 by using the harmonic hall voltage response measurements. Therefore, the room-temperature SOT-driven magnetization switching in a 2D vdW ferromagnet $Fe_3GaTe_2$ could pave the way towards the development of 2D vdW spin-based memory, logic and neuromorphic computing.

## 2. Results and Discussion

### 2.1 Characterizations of $Fe_3GaTe_2$/Pt bilayer

**Figure 1a** schematically shows the spin current in Pt generated by the spin Hall effect,[25–27] which is injected into the adjacent $Fe_3GaTe_2$ layer and exerts SOTs on its magnetization. The high-quality vdW ferromagnetic $Fe_3GaTe_2$ crystal is grown by the self-flux method and reveals a hexagonal structure with a space group $P6_3$/mmc (No. 194).[13] The highly crystalline structure of the layered $Fe_3GaTe_2$ is evident in the representative cross-sectional high-angle annular dark field scanning transmission electron microscope (HAADF-STEM) image of the exfoliated $Fe_3GaTe_2$ nanoflake (see Figure S1). Each layer of $Fe_3GaTe_2$ is constituted by the alternately arranged Te-Fe-Ga (Fe)-Fe-Te atomic planes, and the thickness of monolayer $Fe_3GaTe_2$ is ~0.78 nm.[13]

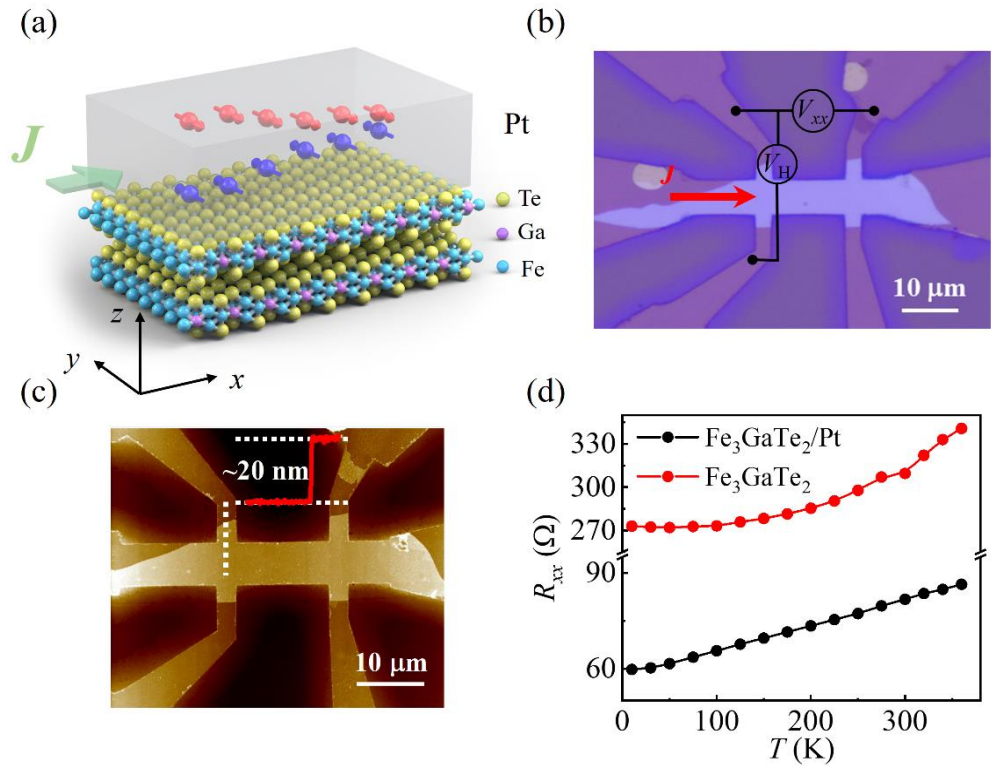

**Figure 1. Schematic diagram and characterizations of 2D $Fe_3GaTe_2$/Pt bilayer.** a) The schematic diagram of the $Fe_3GaTe_2$/Pt bilayer device. The applied electrical current, $J$ (green arrow), in the Pt layer generates a spin current, which injected into the adjacent $Fe_3GaTe_2$ layer. The accumulated spins at the bottom (top) Pt surface are indicated by blue (red) arrows. b) A typical optical image of the $Fe_3GaTe_2$/Pt Hall device. c) The AFM image of the $Fe_3GaTe_2$ and the height profile along the white dashed line,

indicating the thickness of the $Fe_3GaTe_2$ is ~20 nm. d) Longitudinal resistance, $R_{xx}$, versus temperature for $Fe_3GaTe_2$/Pt bilayer and $Fe_3GaTe_2$ only.

In this study, we have fabricated two devices, denoted as devices A and B (the detailed fabrication processes see Experimental Section and Figure S2). A typical optical image of the $Fe_3GaTe_2$/Pt Hall device (device A) is shown in **Figure 1b**. The corresponding atomic force microscopy (AFM) measurements indicate the $Fe_3GaTe_2$ layer thickness is ~20 nm (**Figure 1c** and Figure S3). Prior to investigating the SOT effect in the bilayer hall devices, we characterized the temperature dependent magnetic properties. The longitudinal resistance, $R_{xx}$, of device A as a function of temperature is shown in **Figure 1d**. The $R_{xx}$ increases with increasing temperature, exhibiting a metallic behavior. After considering the resistance of Pt (Figure S4), the resistance of $Fe_3GaTe_2$ can be calculated by deducting the contribution of the Pt layer in the whole device by using parallel resistance model. Then, the extracted $R_{xx}$ of $Fe_3GaTe_2$ increases with increasing the temperature, which is consistent with the previous studies.[13,28]

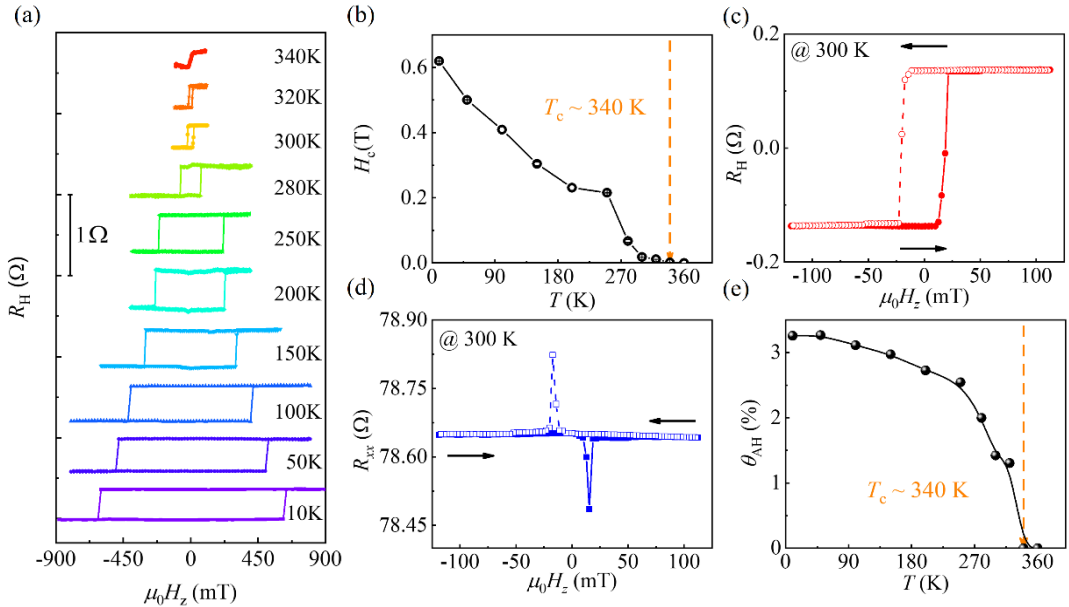

**Figure 2. Magnetic properties of the 2D $Fe_3GaTe_2$/Pt device.** a) Hall resistance, $R_H$, as a function of external perpendicular magnetic field, $\mu_0 H_z$, at different temperatures. As the temperature decreases from 200 K to 10 K, the $R_H$ gradually decreases, which can be attributed to the shunting effect of Pt layer. b) Coercive field, $H_c$, as a function of temperature. c) $R_H$ and d) $R_{xx}$ versus $\mu_0 H_z$ at 300 K. Arrows indicate the field-sweep direction. e) Temperature dependence of anomalous Hall angle, $\theta_{AH}$. The Curie temperature $T_c$ of $Fe_3GaTe_2$ is determained to be about 340 K.

Subsequently, we measured the hysteresis loops of Hall resistance, $R_H$, for different temperatures ranging from 10 K to 360 K (**Figure 2a**). The perfect square-shaped magnetic hysteresis loop holds even up to 340 K, indicating the strong PMA of Fe$_3$GaTe$_2$. Owing to the increasing thermal fluctuations, the coercive field, $H_c$, decreases monotonically with temperature and eventually disappears above $T_c$ ~340 K (**Figure 2b**). The enlarged hysteresis loop of $R_H$ at room temperature (300 K) is shown in **Figure 2c**, in which the coercivity field is about 18 mT. **Figure 2d** shows the magnetic field dependence of the $R_{xx}$ at 300 K. The $R_{xx}$ shows an odd-function response, with a dip (peak) around the positive (negative) switching magnetic field, which is ascribed to the propagation of a domain wall crossing voltage probes.[29,30] Moreover, the temperature dependences of anomalous Hall conductivity, $\sigma_{AH}$, and anomalous hall angle, $\theta_{AH}$, has been calculated (Figure S5 and **Figure 2e**). The $\theta_{AH}$ is expressed as $\theta_{AH} = \sigma_{AH}/\sigma^{FGT}$, where $\sigma^{FGT}$ is the longitudinal conductivity of Fe$_3$GaTe$_2$ (details see Supplementary Note 5). The above-room-temperature $T_c$ and robust PMA in Fe$_3$GaTe$_2$ make it possible to investigate the SOT effect in Fe$_3$GaTe$_2$/Pt device at room temperature.

2.2 Room-temperature SOT switching in 2D Fe$_3$GaTe$_2$/Pt bilayer

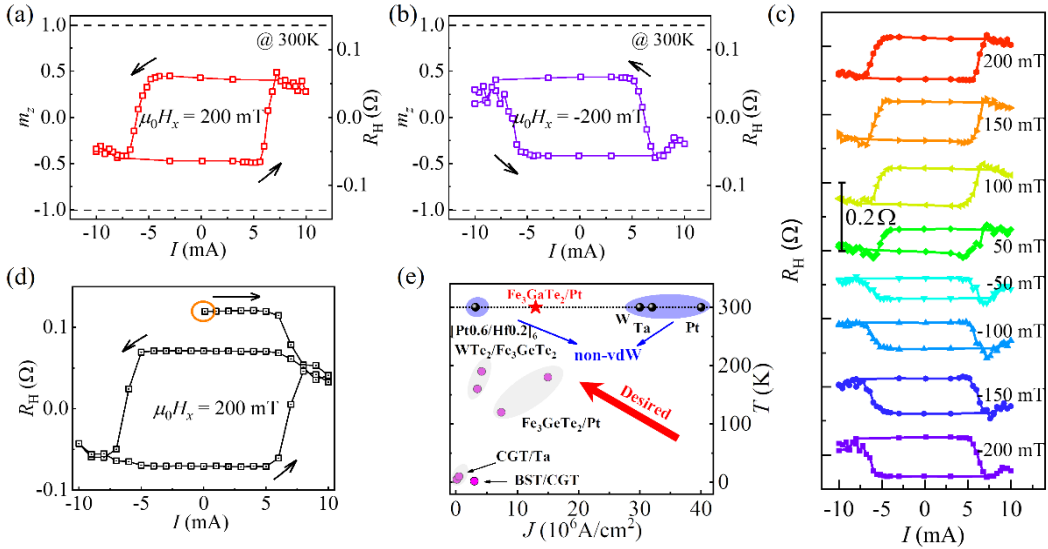

**Figure 3. Room-temperature SOT switching in 2D Fe$_3$GaTe$_2$/Pt bilayer device.** Current-driven perpendicular magnetization switching for a) 200 mT and b) -200 mT in-plane magnetic field at 300 K. The horizontal dashed lines represent the states of saturated magnetization of Fe$_3$GaTe$_2$. Arrows indicate the current-sweep direction. c)

SOT-driven magnetization switching in $Fe_3GaTe_2$ under different in-plane magnetic fields. d) Current-induced perpendicular magnetization switching under 200 mT in-plane magnetic field. The initial state of $Fe_3GaTe_2$ is saturated to the up-state, which is denoted by an orange circle. Arrows indicate the current-sweep direction. e) Comparison of the working temperature and critical current density of our $Fe_3GaTe_2$/Pt bilayer devices with some reported state-of-art SOT devices.

We subsequently examine the possible current-induced magnetization switching in 2D vdW $Fe_3GaTe_2$. With assistance of an in-plane magnetic field $H_x = \pm200$ mT, a 1-ms direct-current (DC) pulse $I$ is first applied along the $x$ direction of the device for the "write" operation, and then a small current of 0.1 mA is applied to "read" the magnetization state of $Fe_3GaTe_2$. As shown in **Figures 3a** and **3b**, the magnetization of $Fe_3GaTe_2$ is switched determinately by the write current and the switching polarity is anticlockwise (clockwise) for the positive (negative) in-plane magnetic field, which is consistent with the sign of the spin-hall angle of Pt ($\theta_{SH} > 0$). This switching behavior suggests that the Pt layer acts as the source of spin currents. Additionally, the critical switching current of 6.3 mA (corresponding current density of $1.3\times10^7 A/cm^2$) is defined by the $dR_H/dI$ verus $I$ curve shown in Figure S6. Moreover, similar switching behaviors can be observed in a wide range of in-plane fields as shown in **Figure 3c** (more details see Figure S7). However, compared to the $R_H$ switched by the external perpendicular magnetic field (denoted as the horizontal dashed lines shown in Figure 3a and 3b), the current-driven $R_H$ is only 40% of that, indicating the current-driven magnetization switching is partial and incomplete. This incomplete switching behavior could be attributed to multiple factors, such as the joule heating effect,[31] the in-plane magnetic field[32] or the pinning effect.[33] As shown in **Figure 3d**, the magnetization of $Fe_3GaTe_2$ begins to decrease as the write current exceeds 6 mA, starting from an initial state which is fully saturated in the +$z$ direction. We observe similar phenomena with different initial states, which indicates the thermal origin (see Figure S8). To prove this, we have analyzed the evolution of the device temperature versus applied write current pulses by tracing the longitudinal resistance of the device (details see supplementary Note 6). It clearly shows that the device temperature reaches the $T_c$ (~340 K) under an 8 mA current pulse and approaches 360 K for a 10 mA current pulse. At such high temperature, the PMA of $Fe_3GaTe_2$ is insufficient to counteract the thermal fluctuations and maintains a single domain state, thereby resulting in the formation of multi-domain states (or unsaturated resistance states). Subsequently, we has investigated the in-plane magnetic field dependence of the switching ratio which shows a non-monotonic dependence (see Figure S9). The highest switching ratio of 0.45 (less than half) suggests the in-plane magnetic field is not the major factor that causes incompleted switching. Additionally, the pinning effect from the Hall voltage probes may exacerbate the incomplete switching since the motion of domain walls is impeded, indicating that the switching ratio can be further improved by shrinking the device size to nano-meter scale.[33] Almost the same room-temperature current-induced magnetization switching behavior has also been observed in another device (as shown in Figure S10), verifying the reproducibility of the devices.

We then evaluate the ability to switch the magnetization of 2D Fe$_3$GaTe$_2$/Pt bilayer by SOT. As evidenced by the asymmetric magnetoresistance observed in **Figure 2d**, the magnetization switching in the Fe$_3$GaTe$_2$/Pt bilayer device is accomplished through domain nucleation and domain wall propagation.[34] We introduce an effective switching efficiency parameter, $\eta$, defined as $\eta = \frac{2e\mu_0 M_S t_{FM}^{eff} H_p}{\hbar J_c}$,[35] where $H_p$ is the domain wall pinning field estimated from the coercive field and $J_c$ is the critical current density. Since the ratio of the current-induced Hall resistance switching to the field-induced Hall resistance switching is about 0.4, it suggests the existance of multi-domain states for the system and only 40% of total magnetic domains are switched from up(down) to down(up) by the SOTs. Thus, we assume the effective thickness of the FM layer, $t_{FM}^{eff} \sim 8$ nm. Taking into account of the Joule heating of the applied current, the actual device temperature is raised to 322 K at the critical current density of 1.3×10$^7$ A/cm$^2$. Assuming the $M_s$ of the Fe$_3$GaTe$_2$ flake in our device is equal to its bulk value,[13] the $M_s$ of 173 emu/cm$^3$ at 322 K can be obtained by following the equation of $M_s(T) = M_s(0 \text{ K}) \times (1 - T/T_c)^\beta$,[36] where $\beta$ is the critical magnetization exponent with a typical value of 0.3 (more details see Supplementary Note 11). By taking $H_p = H_c \sim 11$ mT at 322 K, $\eta \sim 0.35$ can be obtained, which is comparable to that achieved with conventional magnetic materials based SOT devices ($\eta \sim 0.23 - 0.48$).[37] To highlight the performance of our Fe$_3$GaTe$_2$/Pt bilayer device, as shown in **Figure 3e**, we compared the working temperature and critical current density of our SOT device with state-of-art results on vdW magnet-based and CoFeB-based SOT devices, including CGT/Ta,[19,20] (Bi$_{0.5}$Sb$_{0.5}$)$_2$Te$_3$ (BST)/CGT,[18] WTe$_2$/Fe$_3$GeTe$_2$,[16,17] Fe$_3$GeTe$_2$/Pt,[14,15] Ta/CoFeB,[9] Pt/CoFeB,[38] W/CoFeB,[39] [Pt$_{0.6}$/Hf$_{0.2}$]$_6$/CoFeB.[40] The robust room-temperature SOT-driven magnetization switching in Fe$_3$GaTe$_2$/Pt bilayer devices is achieved using a relatively low current density, implying a significant advancement in the field of vdW magnet-based device applications.

## 2.3 The spin-orbit torque efficiency of 2D Fe$_3$GaTe$_2$/Pt bilayer

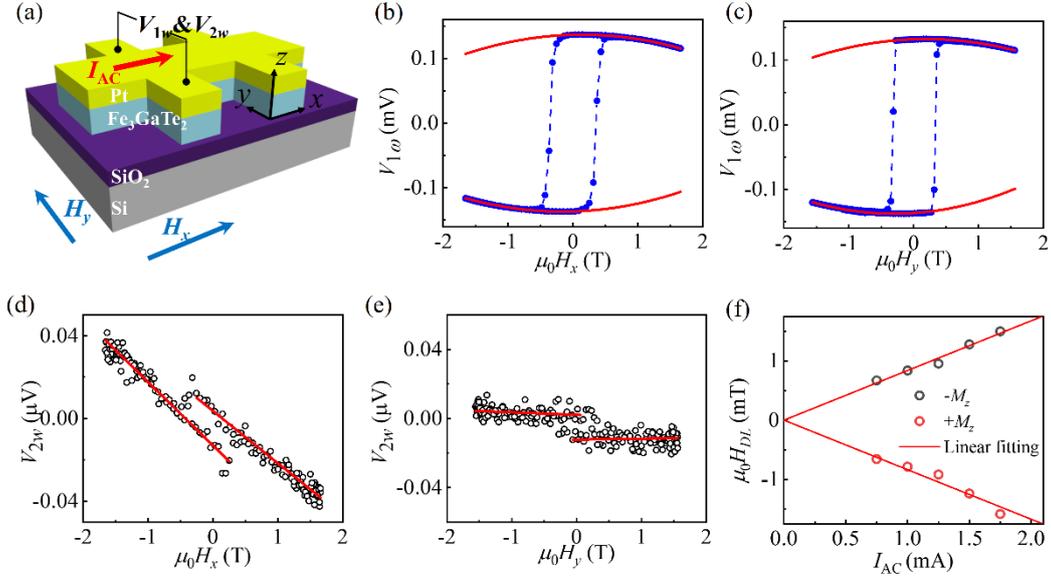

**Figure 4. Characterization of the current-induced SOT effective fields.** a) Geometry and coordinate for the HHVR measurement. First harmonic voltages ($V_{1w}$) as functions of b) longitudinal and c) transverse field. The corresponding second harmonic voltages ($V_{2w}$) as functions of d) longitudinal and e) transverse field. f) Plots of the damping-like field as a function of the alternating current, $I_{AC}$, where the black and red circles are data points measured at different magnetic moment (-$M_z$ and +$M_z$) respectively. The solid lines represent the linear fitting results with zero intercept.

To further quantify the efficiency of SOTs, we employed the harmonic hall voltage response (HHVR) to investigate the SOT effective fields in device A. The experimental set-up and coordiante are shown in **Figure 4a**.[41,42] The small alternating current, $I_{AC}$ (with a frequency of 1327 Hz), is applied under an in-plane magnetic field along the longitudinal (transverse) direction to measure the damping-like (field-like) torque. The representative results of first harmonic voltage ($V_{1w}$) and second harmonic voltage ($V_{2w}$) as functions of the in-plane magnetic field are exhibited in **Figures 4b, 4d** for longitudinal and **Figures 4c, 4e** for transverse directions, respectively. We then extracted the spin-orbit torque effective fields, damping- and field-like components of SOTs, denoted as $H_{DL}$ and $H_{FL}$, by fitting the formula of $H_{DL(FL)} = -2\frac{\partial V_{2\omega}}{\partial H_{x(y)}} \Big/ \frac{\partial^2 V_{1\omega}}{\partial^2 H_{x(y)}}$.[43] The current dependence of $H_{DL}$ is shown in **Figure 4f**, from which the conversion coefficient of ~4.1 mT/(10$^7$ A·cm$^{-2}$) can be obtained by linear fitting data. Noteworthily, the parallelism of the second harmonic voltage in **Figure 4e** between different magnetic moment ($+M_z$ and $-M_z$) suggests a negligible $H_{FL}$. The thermal effect in Fe$_3$GaTe$_2$/Pt bilayer is also neglected for the harmonic measurements due to the small alternating current. Finally, the SOT efficiency, $\xi_{DL}$, of Fe$_3$GaTe$_2$/Pt bilayer device can be derived by using the formula of $\xi_{DL} = \left(\frac{2e}{\hbar}\right)\mu_0 M_s t_{FM}^{eff} H_{DL}/J$,[44,45] where $e$ is the electron charge, $\hbar$ is the reduced Plank constant, $J$ is the current density, and the $t_{FM}^{eff}$ and $M_s$ = 220 emu/cm$^3$ are the effective thickness and saturated magnetization of Fe$_3$GaTe$_2$ layer at 300 K, respectively. The damping-like SOT efficiency is determined to be $\xi_{DL} \sim 0.22$, which is larger than other Pt-based HM/3D FM devices ($\xi_{DL} \sim 0.1$)[37] and comparable to the results of Fe$_3$GeTe$_2$/Pt devices.[15,46] Note that the high damping-like SOT efficiency is also responsible for the low critical current density during the current-induced magnetization switch process.

## 3. Conclusion and outlook

In summary, we have fabricated 2D Fe$_3$GaTe$_2$/Pt bilayer Hall devices by depositing 6-nm Pt onto the few-layered Fe$_3$GaTe$_2$, and the Curie temperature of the device is determined to be 340 K. The robust SOT-driven perpendicular magnetization switching in the 2D Fe$_3$GaTe$_2$ at room temperature have been achieved with a relatively low critical current density of $1.3 \times 10^7$ A/cm$^2$. The low switching current density suggests a high SOT efficiency of our devices which is further determined to be $\xi_{DL} \sim 0.22$. Our results exceed some reported state-of-art SOT devices, and the reproducibility of these findings has been confirmed in another similar device. However, an in-plane magnetic field is still required to break the in-plane symmetry for deterministic switching in our SOT devices, which is challenging for device integration. Specially, the novel 2D low-crystal symmetry quantum material such as WTe$_2$ can generate the unconventional out-of-plane anti-damping SOT, which could be utilized to achieve the deterministic switch in Fe$_3$GaTe$_2$ at room temperature without an external magnetic field.[16,47–49] In conclusion, our results of SOT-driven magnetization switching based on a 2D vdW magnet at room temperature not only paves way towards vdW-based non-volatile memory technologies devices but also offers new opportunities for spin-based logics and neuromorphic computing.

## 4. Experimental Section

**Device fabrication**

To fabricate standard Hall bar devices, $Fe_3GaTe_2$ flakes were firstly exfoliated onto polydimethylsiloxane (PDMS) stamps by adhesive tape. The selected $Fe_3GaTe_2$ flake with appropriate thickness and shape was transferred onto the pre-patterned 300-nm-thick $SiO_2$/Si substrate under an optical microscope. The prepatterned electrodes Cr/Au (5/45 nm) were prepared on a $SiO_2$ substrate by a standard photolithography process. The entire exfoliation and transfer process were conducted in the $N_2$-filled glove box ($H_2O$, $O_2$ < 0.1ppm) to avoid oxidation. To prepare $Fe_3GaTe_2$/Pt heterostructure, the $SiO_2$/Si substrates with exfoliated FGT flakes were transferred immediately into the loadlock of our sputtering system. Then a 6 nm layer of Pt was deposited on the $Fe_3GaTe_2$ surface, and the obtained $Fe_3GaTe_2$/Pt bilayer is air-stable, which is conducive to the subsequent manufacturing process. Once removed from the sputtering chamber, the $Fe_3GaTe_2$/Pt bilayer were patterned into Hall bar devices with channel width of 6 μm using a standard electron-beam lithography and Ar ion etching technique.

**Characterization**

The anomalous hall effect measurements (AHE) were conducted using a Model CRX-VF Cryogenic Probe Station (Lake Shore Cryotronics, Inc.) with a ± 2.25 T of out-of-plane perpendicular magnetic field. The AHE and longitudinal resistance were measured in the temperature range of 10 to 360 K, a small reading current $I_{read}$ = 0.1 mA was applied during the test. The SOT switching and HHVR measurements were carried out at room temperature. For the HHVR measurement, we used the Keithley 6221 to apply a low-frequency alternation current $I(t) = I_{AC}\cos(2\pi ft)$ (with $f$ = 1327Hz). The first- and second-harmonic voltages were measured using double Stanford SR830 Lock-in amplifiers. For current-induced magnetization switching measurements, a pulsed electrical current with a duration of 1 ms was applied, and subsequently, $I_{read}$ was applied to read out the Hall voltage after waiting for 1 s. The cross-sectional scanning transmission electron microscope (STEM) samples were prepared by using a focused ion beam (FIB, FEI Helios NanoLab 600i), and were characterized by using an aberration-corrected STEM (JEM-ARM200F NEOARM, JEOL Ltd.) under high-angle annular dark field (HAADF) and annular bright-field (ABF) imaging modes.

**Supporting Information**

Supporting Information is available from the Wiley Online Library or from the author.

## Data availability

The data that support the findings of this study are available from the corresponding author upon reasonable request.

## Acknowledgments

This work was financially supported by the National Key Research and Development Program of China (Grant Nos. 2022YFA1405100 and 2022YFE0134600), the Beijing Natural Science Foundation Key Program (Grant Nos. Z190007 and Z220005), the National Natural Science Foundation of China (Grant Nos. 12241405, 12174384 and 52272152), and the Strategic Priority Research Program of Chinese Academy of Sciences (Grant Nos. XDB44000000 and XDB28000000).

## Conflict of interest

The authors declare no conflicts of interest.